# Time in Quantum Theory[*]

H. D. Zeh
www.zeh-hd.de

**I.** In general, time is used in quantum theory as an external ('classical') concept. So it is assumed, as in classical physics, to exist as a controller of all motion – either as absolute time or in the form of proper times defined by a classical spacetime metric. In the latter case it is applicable to local quantum systems along their world lines. According to this assumption, time can be read from appropriate classical or quasi-classical 'clocks'.

This conception has to be revised only when relativistic gravity, which regards the spacetime metric as a dynamical object, is itself quantized [1] – as required for consistency (see **IV**). The thereby aspired quantization of time does not necessarily lead to a discretization of time – just as the quantization of free motion does not require a discretization of space or position. On the other hand, the introduction of a fundamental gravitational constant in addition to Planck's constant and the speed of light leads to a natural 'Planck time' unit, corresponding to $5.40 \; 10^{-44}$ *sec*. This may signal the need for an entirely novel conceptual framework – to be based on as yet missing empirical evidence. A formal (canonical) quantization of time would also be required in non-relativistic *Machian* ('relational') dynamical theories [4], which consistently replace the concept of time by some reference motion. If quantum theory is universally valid, *all* dynamical processes (including those that may serve as clocks or definers of time) must in principle be affected by quantum theory. What does this mean for the notion of time?

Historically, the dynamics of quantum systems seemed to consist of individually undetermined stochastic 'quantum jumps' between otherwise 'stationary' states (energy eigenstates) – see [2] for an early review of the formalism and the attempt of an interpretation. Such stochastic events are observed in quantum measurements, in particular. For this reason, von Neumann [3] referred to the time-dependent →Schrödinger equation as a 'second intervention', since Schrödinger had invented it solely to describe consequences of time-dependent external 'perturbations' on a quantum system. Note, however, that atomic clocks are *not* based on any stochastic quantum events, even though they have to be designed as open systems in order to allow their permanent reading (representing 'measurements' of the clock – see **IV**).

In a consistent →Schrödinger picture, *all* dynamics is described as a time dependence of the quantum *states*, while the observables are fixed formal kinematical concepts (see also Sect. 2.2 of [5]). The time dependence according to the Schrödinger equation can be completely understood as an interference phenomenon between different stationary states $|m>$, which possess individually meaningless phase factors $\exp(i\omega_m t)$. Their →superpositions are able to describe time-dependent quantum states $|\alpha(t)>$ in the form

$$|\alpha(t)> := \int dq \; \psi_\alpha(q,t)|q> = \Sigma_m c_m \exp(i\omega_m t)|m> \;\;.$$

The wave function $\psi_\alpha(q,t)$ is here used to define the time-dependent state $|\alpha(t)>$ in abstract →Hilbert space. The Hilbert space basis $|q>$ diagonalizes an appropriate observable $Q$. The time dependence of a quantum state is in fact meaningful only *relative* to such a fixed basis, as demonstrated by means of the wave function in the above definition.

In non-relativistic quantum mechanics, the time parameter $t$ that appears in the Schrödinger wave function $\psi(q,t)$ is identified with Newton's absolute time. So it is presumed to

---

exist regardless of how or whether it is measured. The letter *q* represents all variables $q_i$ (*i=1…I*) that span the required configuration space. The special case of a point mass, where *q* ≡ *x,y,z* corresponds to a single space point, has often led to confusion of the wave function with a time-dependent spatial *field* (relativistically a field on spacetime). In *N*-particle mechanics, for example, the configuration space variables *q* are equivalent to *N* space points (that is, *I=3N* variables). In quantum field theory, the amplitudes of all fields *Φ(x,y,z)* at all space points even form a continuum. These variables are thus distinguished from one another by their spatial arguments, which now assume the role of 'indices' to *Φ*, just as *i* for the variables $q_i$ [6]. Therefore both, space and time, are in this approach assumed to be absolutely defined classical preconditions for kinematics and dynamics – even though they appear in the formalism in different ways.

If the variables *q* are field amplitudes, the canonical quantization of n fields, for example, leads to a time-dependent wave functional *Ψ[Φ₁(r),…,Φₙ(r),t]*, rather than to *n* field operators on spacetime. This conclusion holds relativistically, too. The corresponding Hilbert space readily includes superpositions of different 'particle' numbers ('occupation numbers'). For bosons, the latter are simply oscillator quantum numbers for the eigenmodes (first postulated by Planck, and later explained by Schrödinger by the numbers of nodes of their wave functions). The ultimate universal Hilbert space basis is hoped to be discovered in a unified field theory.

Schrödinger's general wave function *ψ(q,t)* may be Fourier transformed with respect to *all* its arguments – in spite of their different interpretations. This transformation defines wave numbers *k* in the *configuration* space and frequencies *ω*. They may be rescaled into canonical momenta (in general different from conventional, that is, spatial momenta) and energies by means of Planck's constant. The Fourier transformation gives rise to a formal 'time operator', *T* := *i∂/∂ω*, that allows one to define a continuous shift operation for frequencies: *U(Δω)* := exp*(iΔωT)*. It does *not* in general transform a solution of the Schrödinger equation into another solution, since this would require a continuous and unbounded energy spectrum. Pairs of Fourier variables are subject to the Fourier theorems,

$$\Delta q \Delta k \geq 1 \quad \text{and} \quad \Delta t \Delta \omega \geq 1 \; ,$$

which apply to all functions *ψ(q,t)* – regardless of the existence of any dynamical law or a Hamiltonian *H*. These 'uncertainty relations' between corresponding variables must have *physical* consequences when applied to solutions of the Schrödinger equation. Those based on the Fourier theorem relating time and frequency are usually interpreted as representing a 'time-energy uncertainty relation' (see [7]). Well known, for example, is the spectral line width required for metastable states. A 'time uncertainty' can also be defined by the finite duration of a preparation or measurement process.

**II.** The situation is somewhat obscured in the →Heisenberg picture. In the algebraic Born-Heisenberg-Jordan quantization procedure, 'observables' were introduced in formal analogy to the classical *dynamical variables*, such as *q(t)* and *p(t)*, while quantum states were *not* regarded as dynamical objects. Observables would assume definite *values* only in appropriate measurements or discrete 'quantum events' (von Neumann's first intervention – historically related to Bohr's quantum jumps between his discrete classical orbits). Time *durations* are then often defined operationally by means of pairs of such events – not according to the Schrödinger dynamics. The latter is here merely regarded as a tool for calculating probabilities for the occurrence of events, which are thereby assumed to represent the only *real* quantum phenomena.

Note that in the Heisenberg picture certain *instantaneous* properties of quantum states appear to represent some hidden time dependence. For example, the kinetic energy operator in the Schrödinger picture (the Lapacean) measures the curvature of the wave function *ψ(q,t)* at given time *t* – not any quantitiy related to motion, such as the classical kinetic energy, while its minimum (achieved for a wave function that nowhere changes sign) is in the Heisenberg picture interpreted as describing 'zero point fluctuations' of the corresponding variables $q$.

This picture has led to much confusion – including the search for a 'time observable' $T$ that would depend on the specific system Hamiltonians $H$ by obeying commutation relations
$$[T,H] = i\hbar ,$$
in analogy to position and momentum observables (see the Introduction to [8] for a review). However, since realistic Hamiltonians possess a ground state, their spectra are bounded from below, and a time operator obeying this commutation relation cannot possess a spectrum represented by the real numbers (as pointed out by Wolfgang Pauli [2]). It may nonetheless be related to time intervals between certain pairs of events that can be measured at a system characterized by the Hamiltonian $H$.

A formal equivalence between the Schrödinger and a Heisenberg picture for the purpose of calculating expectation values of measurement results is known to hold for isolated, unitarily evolving systems (which are exceptions in reality). For *asymptotically* isolated objects participating in a scattering process one may use the interaction picture, where part of the Hamiltonian dynamics is absorbed into the observables characterizing asymptotic states. This includes the 'dressing' of quantum fields. However, *macroscopic* systems always form open systems; they never become isolated, even when dressed. Such systems may approximately obey effective non-unitary dynamics (master equations). In principle, this dynamics has to be derived from the unitary (Schrödinger) evolution of an entangled *global* quantum state, that would include all 'external interventions'. Under realistic assumptions this leads to permanently growing →entanglement with the environment – locally observed as →decoherence [5].

This extremely fast and in practice irreversible process describes a *dislocalization* of quantum superpositions. It thereby mimics quantum jumps (events): components which represent different macroscopic properties (such as different pointer positions or different registration times of a detector) are almost immediately dynamically decoupled from one another – though none of them is selected as the *only* existing one. Pauli, when arguing in terms of the Heisenberg picture, regarded such events as occuring 'outside the laws of nature', since they withstood all attempts of a local dynamical description. In the global Schrödinger picture, the time-asymmetry of this dynamical decoupling of components ('branching') can be explained in terms of the time-symmetric dynamics by means of an appropriate initial condition for the wave function of the universe – the same condition that may also explain thermodynamical and related time asymmetries ('arrows of time') [9]. In essence, this initial condition requires that non-local entanglement did not yet exist just after the big bang, and therefore has to *form* dynamically ('causally'). The resulting asymmetry in time may give rise to the impression of a *direction of time*.

**III.** In quantum field theory, a Schrödinger equation that controls the dynamics of the field functionals may well be relativistic – containing only local interactions with respect to the space-dependent field variables (in this way facilitating the concept of a *Hamiltonian density* in space). A wave function(al) obeying a relativistic Schrödinger equation never propagates faster than light with respect to the underlying presumed absolute spacetime. Recent reports of apparently observed superluminal phenomena were either based on inappropriate clocks, or on questionable interpretations of the wave function. For example, the exact energy eigenstate of a particle, bound to an attractive potential in a state of negative energy $E = -|E|$, would extend to spatial infinity according to $\exp(-\sqrt{|E|}r)$ outside the range of the potential. It has therefore been claimed to be able in principle to cause effects at an arbitrary distance within any finite time [10]. However, if the wave function of the bound system forms dynamically (according to the Schrödinger equation rather than by quantum jumps), it can only subluminally *approach* the exact eigenstate with its infinite exponential tail. This time-dependence requires a minimum energy spread that is in accord with the time-frequency Fourier theorem. Similar arguments hold relativistically also for particle number eigenstates, which cannot have sharp spatial boundaries because of Casimir type effects (in principle observable for moving mirrors); all bounded systems must relativistically be in superpositions of diffent particle numbers.

In the theory of relativity, proper times assume the role of Newton's absolute time for all *local* systems, that is, for those approximately following world lines in spacetime. However, quantum states are generically nonlocal, and they do not consist of or define local subsystem states. One may then introduce auxiliary time coordinates (arbitrary spacetime foliations) in order to define the dynamics of *global* states on these artificial 'simultaneities'. A Hamiltonian (albeit of very complex form – in general including a whole field of Coriolis-type forces with effective 'particle' creation and annihilation terms) would nonetheless *exist* in this case. As these artificial simultaneities may be assumed to propagate just locally, one speaks of 'many-fingered time'. Dynamical evolution in quantum theory is in general *locally non-unitary* (to be described by a master equation) because of the generic nonlocal entanglement contained in the unitarily evolving global quantum state. Unitary evolution may therefore be confirmed only in exceptional, quasi-isolated (microscopic) systems.

**IV.** According to Mach's ideas, no concept of absolute time should be required or meaningful. Any time concept could then be replaced by simultaneity *relations* between trajectories of different variables (including appropriate clocks) – see [4] and Chap. 1 of [9]. *Classically,* timeless trajectories $q(\lambda)$, where $\lambda$ is an arbitrary and physically meaningless parameter, are still defined. Mach's principle requires only that the fundamental dynamical laws are invariant under reparametrizations of $\lambda$. *In quantum theory,* the wave function cannot even depend on such a time-ordering parameter, since there are no trajectories any more that could be parametrized. This excludes even dynamical successions of spatial geometries (the dynamical states of general relativity), which would form a foliation of spacetime. On the other hand, any appropriate variable $q_0$ that is among the arguments of a time-less wave function $\psi(q)$ may be regarded as a more or less appropriate global physical clock. According to the superposition principle, *superpositions* of different values $q_0$ – that is, of different 'physical times' – would then have to exist as real physical states (just as the superpositions of different values of any physical variable).

In conventional quantum mechanics, superpositions of different times of an event are well known. For example, a coherently decaying metastable state (that can be experimentally confirmed to exist by means of interference in the case of decay fragments only weakly interacting with their environment) is a superposition of different decay times. Similarly, the quantum state for a single variable $x$ and a clock variable $u$, say, would have to be described by a wave function $\psi(x,u)$. This means that the *classical* dependence of $x$ on clock time $u$, defined by their time-less trajectory $x(u)$, is replaced by the less stringent *entanglement* between $x$ and $u$ that is defined by such a wave function [11]. The clock variable $u$ becomes quasi-classical only if it is robust under environmental decoherence, such that superpositions of different times $u$ always remain dislocalized (locally inaccessible). The same conclusion holds for the mentioned superposition of different decay times if its corresponding partial waves (wave packets forming thin spherical shells in space unless reflected somewhere) are decohered from one another.

Atomic clocks, in particular, are based on the time-dependent superposition of two close atomic energy eigenstates (defining 'beats'). These oscillating states would immediately decohere whenever they were measured (read). Therefore, they have to be dynamically correlated with the coherent state of a maser field that is in resonance with them. This time-dependent coherent state is known to be 'robust' against decoherence – including genuine measurements [12]. So it permits the construction of a quasi-classical atomic clock that can be read. Exactly classical clocks would be in conflict with the uncertainty relations between position and momentum of their 'hands'.

The above-described consequences of Mach's principle with respect to time do indeed apply *in general relativity* to a closed universe. Spatial geometries on a time-like foliation of spacetime, which would classically determine all proper times [13], are now among the *dynamical variables q* (arguments of the wave function) – similar to the mentioned clock variable $u$. Moreover, *material* clocks intended to 'measure' these proper times within a given precision would have to possess a minimum mass in order to comply with the uncertainty relations [14], while this mass must then in turn disturb the spacetime metric.

A time *coordinate t* in general relativity is a physically meaningless parameter (such as $\lambda$ – not $u$ – in the above examples). Invariance of the theory under reparametrization, $t \rightarrow f(t)$, requires a 'Hamiltonian constraint': $H = 0$ [1,15]. In its quantum mechanical form, $H\Psi = 0$, this leads to the trivial Schrödinger dynamics $\partial\Psi/\partial t = 0$, where $\Psi$ is now a wave functional on a configuration space consisting of spatial geometries and matter fields. As this consequence seems to remain valid for all unified theories that contain →quantum gravity, one has to conclude that *there is no time* on a fundamental level; all dynamics is encoded in the static entanglement described by $\Psi$. Surprisingly, though, the time-less *Wheeler-DeWitt equation* [1],

$$H\Psi = 0 \ ,$$

(also called an Einstein-Schrödinger equation) becomes hyperbolic for Friedmann type universes – similar to a relativistic wave equation on spacetime (see Sect. 2.1 of [9]). This allows one to formulate a complete boundary condition for $\Psi$ in the form of an 'intrinsic initial condition' [16]. It requires $\Psi$ and its first derivative to be given on a 'time-like' hypersurface, defined according to the hyperbolic form of the kinetic energy operator contained in $H$ (now a d'Alembertian), in this universal configuration space (DeWitt's 'superspace'). For example, such initial data can be freely chosen at a small value of the expansion parameter $a$ of the universe. A low-entropy condition at $a \rightarrow 0$ then leads to an 'intrinsic arrow of time': total entropy on time-like hypersurfaces must grow (for statistical reasons) as a function of the size of the universe – regardless of any external concept of time.

Quasi-classical time can here only be recovered within the validity of a Born-Oppenheimer approximation with respect to the square root of the inverse Planck mass [15], while spatial geometry, which defines all fundamental physical clocks, is strongly entangled with, and thus decohered by, matter [17]. In analogy to the coherent set of apparent light rays that approximately describe the propagation of *one* extended light wave in space in the limit of short wave lengths (geometric optics), quasi-classical times are defined approximately, but *separately* for all quasi-trajectories in superspace. Each of them then defines a dynamically autonomous quasi-classical world (an 'Everett branch' of the global wave function in unitary description) – including a specific quasi-classical spacetime. As 'Schrödinger cat' states evolve abundantly from microscopic superpositions in measurement-type interactions, there cannot be just one quasi-classical world (analogous to just one light ray) according to the Schrödinger dynamics. Material clocks, such as atomic clocks, require further (usually not quite as strong) decoherence to become quasi-classical.